\pdfoutput=1
\documentclass[a4paper]{article}

\usepackage{INTERSPEECH2018}

\usepackage{amsmath,amssymb}
\usepackage{graphicx}
\usepackage{multirow} 
\graphicspath{{grf/}}

\newcommand{\X}{\ensuremath{\mathbf{X}}}

\newcommand{\f}{\ensuremath{\mathbf{f}}}

\newcommand{\h}{\ensuremath{\mathbf{h}}}

\newcommand{\w}{\ensuremath{\mathbf{w}}}
\newcommand{\x}{\ensuremath{\mathbf{x}}}
\newcommand{\y}{\ensuremath{\mathbf{y}}}

\newcommand{\bbR}{\ensuremath{\mathbb{R}}}

\newcommand{\calL}{\ensuremath{\mathcal{L}}}

\newcommand{\caja}[4][1]{{%
    \renewcommand{\arraystretch}{#1}%
    \begin{tabular}[#2]{@{}#3@{}}%
      #4%
    \end{tabular}%
    }}

{%
\begin{list}{#1}{
\vspace{-\topsep}
\vspace{-\partopsep}
\setlength{\itemindent}{0cm}
\setlength{\rightmargin}{0cm}
\setlength{\listparindent}{0cm}
\settowidth{\labelwidth}{#1}
\setlength{\leftmargin}{\labelwidth}
\addtolength{\leftmargin}{\labelsep}
\setlength{\itemsep}{0cm}
}%
}%
{%
\end{list}
\vspace{-\topsep}
\vspace{-\partopsep}
}

{\begin{enumerate}%
}%
{\end{enumerate}}

\newcommand{\ie}{i.e.\@}

\newcommand{\rbr}[1]{\left(#1\right)}
\newcommand{\cbr}[1]{\left\{#1\right\}}

\usepackage{color}
\usepackage{colortbl}
\definecolor{light-gray}{gray}{0.8}

\title{A simple model for detection of rare sound events}
\name{Weiran Wang,  Chieh-chi Kao,  Chao Wang}
\address{Amazon Alexa  \\ 101 Main St, Cambridge, MA 02142, USA}
\email{\{weiranw,chiehchi,wngcha\}@amazon.com}

\begin{document}
\maketitle

\begin{abstract}
We propose a simple recurrent model for detecting rare sound events,
when the time boundaries of events are available for training. 
Our model optimizes the combination of an utterance-level loss, which
classifies whether an event occurs in an utterance, and a frame-level loss, which
classifies whether each frame corresponds to the event when it does occur. 
The two losses make use of a shared vectorial representation the event,
and are connected by an attention mechanism. 
We demonstrate our model on Task 2 of the DCASE 2017 challenge, and achieve
competitive performance.
\end{abstract}

\section{Introduction}
\label{s:intro}

The task of detecting rare sound events from audio has drawn much recent attention, due to its wide applicability for acoustic scene understanding and audio security surveillance. The goal of this task is to classify if certain type of event occurs in an audio segment, and when it does occur, detect also the time boundaries (onset and offset) of the event instance.  

The task 2 of DCASE 2017 challenge provides an ideal testbed for detection algorithms~\cite{Mesaro_17a}. The data set consists of isolated sound events for three target classes (baby crying, glass breaking, and gun shot) embedded in various everyday acoustic scenes as background. Each utterance contains at most one instance of the event type, and the data generation process provides temporal position of the event which can be used for modeling. 

The most direct solution to this problem is perhaps to model the hypothesis space of segments, and to predict if each segment corresponds to the time span of the event of interest. This approach was adopted by~\cite{Wang_17d} and~\cite{Kao_18a}, whose model architecture heavily drew inspirations from the region proposal networks~\cite{Ren_15a} developed in the computer vision community. There are a large number of hyper-parameters in such models, which requires much human guidance in tuning. More importantly, this approach is generally slow to train and test, due to the large number of segments to be tested.
 
Another straight-forward approach to this task is to generate reference labels for each frame indicating if the frame correspond to the event, and then train a classifier to predict the binary frame label. This was indeed the approach taken by many participants of the challenge (e.g., ~\cite{Lim_17a, Cakir_17a}). The disadvantage of this approach is that it does not directly provide an utterance-level prediction (if an event occurs at all), and thus requires heuristics to aggregate frame-level evidence for that. It is the motivation of our work to solve this issue.

We propose a simple model for detecting rare sound events without aggregation heuristics for utterance-level prediction. Our learning objective combines a frame-level loss similar to the abovementioned approach, with an utterance-level loss that automatically collects the frame-level evidence.  The two losses share a single classifier which can be seen as the vectorial representation of the event, and they are connected by an attention mechanism. 
Additionally, we use multiple layers of recurrent neural networks (RNNs) for feature extraction from the raw features, and we propose an RNN-based multi-resolution architecture that consistently improve over the standard multi-layer bi-directional RNNs architectures for our task. 
In the rest of this paper, we discuss our learning objective in Section~\ref{s:model},  introduce the multi-resolution architecture in Section~\ref{s:multires}, demonstrate them on the DCASE challenge in Section~\ref{s:expt}, and provide concluding remarks in Section~\ref{s:conclusion}.

\section{Our model}
\label{s:model}

Denote an input utterance by $\X = [\x_1, \dots, \x_T]$ where $\x_i \in
\bbR^d$ contains the audio features for the $i$-th frame. For our task
(detecting a single event at a time), we are given the binary utterance
label $y$ which indicates if an event 
occurs ($y=1$) or not ($y=0$). If $y=1$, we have additionally the onset
and offset time of the event, or equivalently frame label $\y = [y_1,
\dots, y_T]$, where $y_t=1$ if the event is on at frame $t$ and $y_t=0$
otherwise. 
Our goal is to make accurate predictions at both the utterance level and the frame level. 

Our model uses a multi-layer RNN architecture $f$ to extract nonlinear features from
$\X$, which yields a new representation 
\begin{align*}
f(\X) = [\h_1,\dots,\h_T] \in \bbR^{h\times T},
\end{align*}
containing temporal information. We also learn a vectorial representation
of the acoustic event by $\w \in \bbR^h$, which serves the purpose of a classifier and will be used in predictions at two levels.

With the standard logistic regression model, we perform per-frame
classfication based on the frame-level representation and the classifier
$\w$: for  $t=1,\dots,T$,
\begin{align*}
p_t := P (y_t=1 | \X) = \frac{1}{1 + \exp \rbr{- \w^\top \h_t}} \in [0,1],
\end{align*}
and we measure the frame-level loss if the event occurs:
\begin{align*}
& \calL_{frame} (\X, \y) = \\ 
& \left\{ \begin{array}{c@{\hspace{1ex}}@{:}@{\hspace{1ex}}l}
\frac{1}{T} \sum_{t=1}^T y_t \log p_t + (1 - y_t) \log (1 - p_t) & y=1 \\
0 & y=0 
\end{array}
\right. .
\end{align*}
Note that we do not calculate the frame loss if no event occurs, even
though one can consider the frame label to be all $0$'s in this case. This
design choice is consistent with the evaluation metric for rare events,
since if we believe no event occurs in an utterance, the onset/offset or the frame labels are meaningless. 

\begin{figure*}
\centering
\includegraphics[width=0.85\linewidth, bb=50 20 800 550, clip]{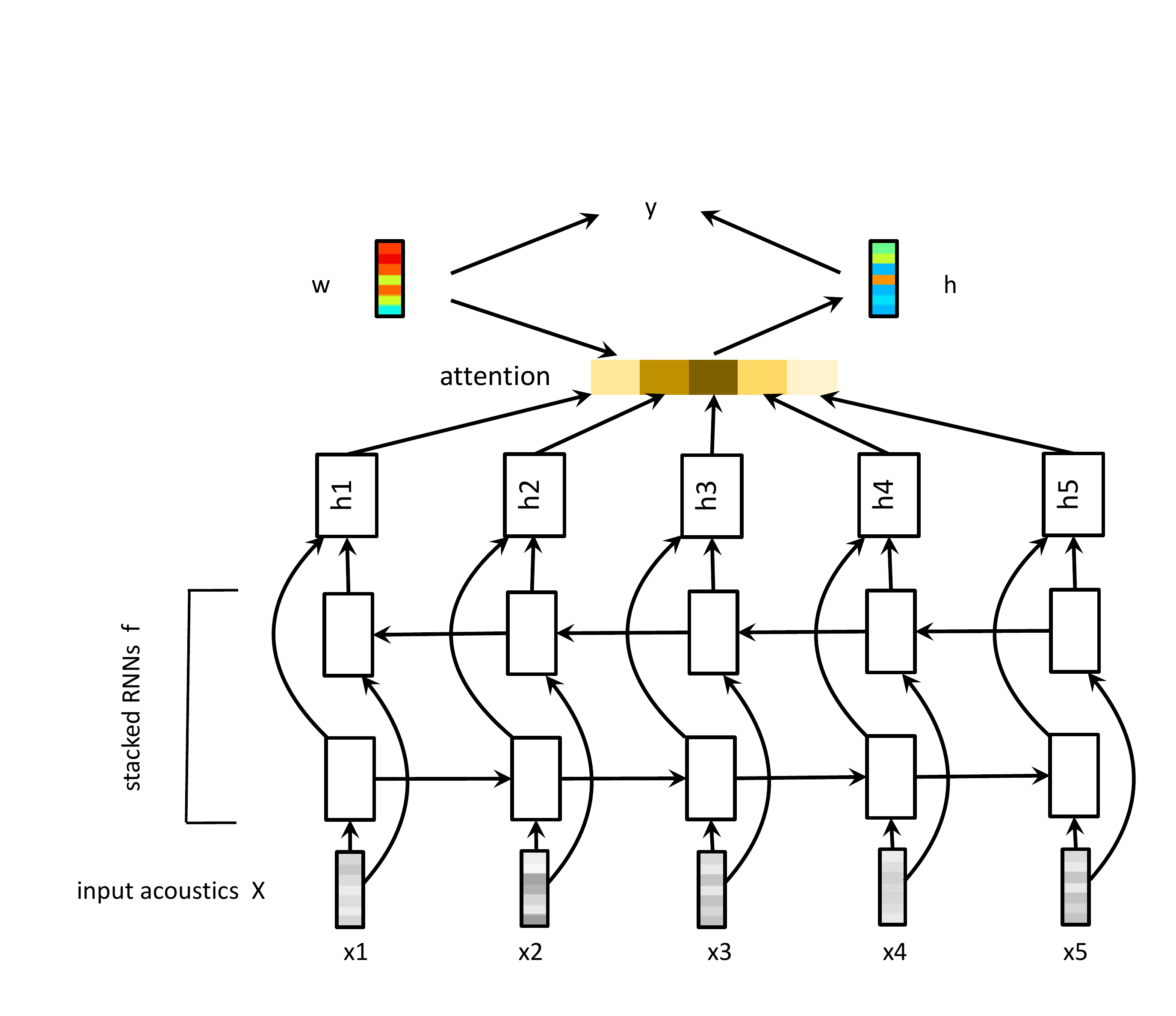}
\vspace*{-3ex}
\caption{Illustration of our RNN-based attention mechanism for rare sound events detection.}
\label{f:model}
\end{figure*}

On the other hand, we make the utterance-level prediction by collecting
evidence at the frame level. Since the above $p_t$'s provide the alignment
between each frame and the target event, we normalize them over the
entire utterance to give the ``attention''~\cite{Bahdan_15a,Bahdan_15b}:
\begin{align*}
a_t = \frac{p_t}{\sum_{t=1}^T p_t}, \qquad t=1,\dots,T,
\end{align*}
and use these attention weights to combine the frame representations to form the utterance representation as
\begin{align*}
\h = \sum_{t=1}^T a_t \h_t.
\end{align*}
We make utterance-level prediction by classifying $\h$ using $\w$:
\begin{align*}
p := P (y=1| \X) = \frac{1}{1 + \exp \rbr{- \w^\top \h}} \in [0,1],
\end{align*}
and define the utterance-level loss based on it:
\begin{align*}
\calL_{utt} (\X, y) = y \log p + (1 - y) \log (1 - p) .
\end{align*}
This loss naturally encourages the attention to be peaked at the event
frames (since they are better aligned with $\w$), and
low at the non-event frames. 

Our final objective function is a weighted combination of the two above losses:
\begin{align*}
\calL (\X, y, \y) = \calL_{utt} (\X, y) + \alpha \cdot \calL_{frame} (\X, \y),
\end{align*}
where $\alpha>0$ is a trade-off parameter. 
During training, we optimize $\calL (\X, y, \y)$ jointly over the
parameters of RNNs $\f$ and the event representation $\w$. 
An illustration of our model is given in Figure~\ref{f:model}.

\subsection{Inference} 
For a test utterance, we first calculate $p$ and
predict that no event occurs if $p \le thres_0$, and in the case of $p >
thres_0$ which indicates that an event occurs, we threshold
$[p_1,\dots,p_T]$ by $thres_1$ to predict if the event occurs at each
frame. For the DCASE challenge task 2, where we need to output the time
boundary for a predicted event (and there is at most one event in each
utterance), we simply return the boundary of the longest connected
component of $1$'s in the thresholded frame prediction. 
We have simply used  $thres_0 = thres_1=0.5$ in our experiments.

\section{Multi-resolution feature extraction}
\label{s:multires}

\begin{figure*}[h]
\centering
\includegraphics[width=0.85\linewidth, bb=0 0 800 580, clip]{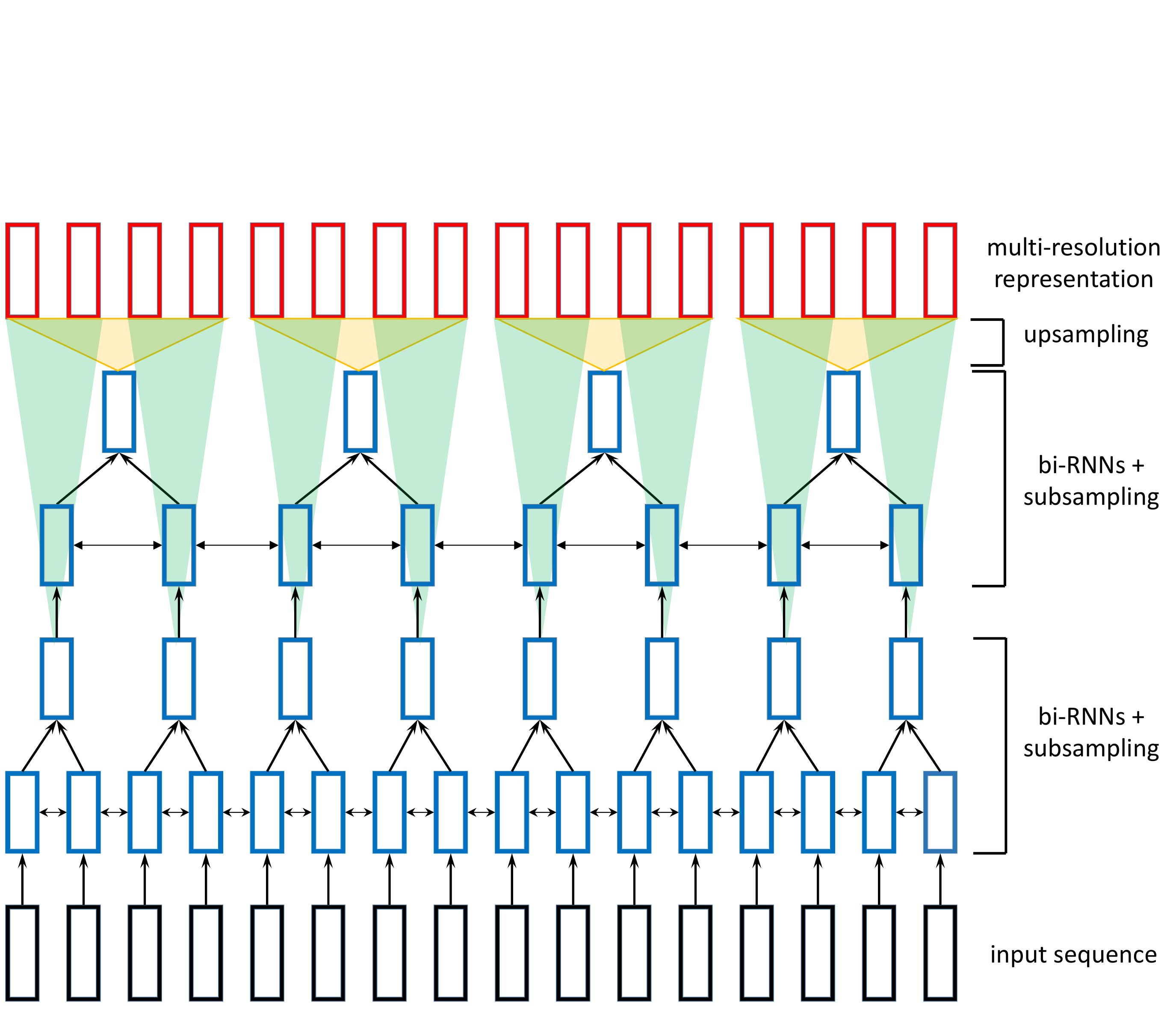}
\vspace*{-2ex}
\caption{RNN-based multi-resolution modeling.}
\label{f:multires}
\end{figure*}

Different instances of the same event type may occur with somewhat different
speeds and durations. To be robust to variations in the time axis, we propose a
multi-resolution feature extraction architecture based on RNNs, as
depicted in Figure~\ref{f:multires}, which will be used as the $f(\X)$ mapping
in our model.

This architecture works as follows. After running each recurrent layer, 
we perform subsampling in the time axis with a rate of 2, \ie, the outputs of 
the RNN cell for two neighboring frames are averaged, and the resulting 
sequence, whose length is half of the input length of this layer, 
is then used as input to the next recurrent layer. In such a way, the
higher recurrent layers effectively view the original utterance at coarser
resolutions (larger time scales), and extract information from increasingly larger context of the input.  

After the last recurrent layer, we would like to obtain a representation
for each of the input frames. This is achieved by upsampling (replicating) the
subsampled output sequences from each recurrent layer, and summing them for
corresponding frames. Therefore, the final frame representation produced
by this architecture takes into account information at different
resolutions. 
We note that the idea of subsampling in deep RNNs architecture is
motivated by that used in speech recognition~\cite{Miao_16a}, and the idea of
connecting lower level features to higher layers is similar
to that of resnet~\cite{He_16a}. 
We have implemented our model in the tensorflow framework~\cite{Abadi_15a}.

\section{Experimental results}
\label{s:expt}

\noindent \textbf{Data generation}
We demonstrate our rare event detection model on the task 2 of DCASE 2017
challenge~\cite{Mesaro_16a} . The task data consist of isolated sound
events for three target events (babycry, glassbreak, gunshot), and recordings of 15 different audio scenes (bus,
cafe, car, etc.) used as background sounds from TUT Acoustic Scenes 2016
dataset~\cite{Mesaro_16b}. 
The synthesizer provided as a part of the DCASE challenge is
used to generate the training set, and the mixing event-to-background
ratios (EBR) are $-6$, $0$ and $6$ dB. 
The generated training set has 5000 or 15000 utterances for each target
class, and each utterance contains either one target class event or no events. 
We use the same development and evaluation set (both of about $500$ utterances) provided by the DCASE challenge.

\noindent \textbf{Feature extraction}
The acoustic features used in this work are log filter bank energies
(LFBEs). The feature extraction operates on mono audio signals sampled at
44.1 kHz. For each 30 seconds audio clip, we extract 64 dimensional LFBEs from frames of 46 ms duration
with shifts of 23 ms. 
 
\noindent \textbf{Evaluation metrics}
The evaluation metrics used for audio event detection in DCASE 2017 are 
event-based error rate (ER) and F1-score. 
These metrics are calculated using onset-only condition with a collar of
500 ms, taking into account insertions, deletions, and substitutions of
events. Details of these metrics can be found in~\cite{Mesaro_16a}.

\subsection{Training with 5K samples}

For each type of event, we first explore different architectures and hyperparameters on 
training sets of 5000 utterances, 2500 of which contain the event.  
This training setup is similar to that of several participants of the
DCASE challenge. 

For the frame-level loss $\calL_{frame}$, instead of summing the 
cross-entropy over all frames in a positive utterance, we only consider 
frames near the event and in particular,  from 50 frames before the onset 
to 50 frames after the offset. In this way, we obtain a balanced set of
frames (100 negative frames and a similar amount of
positive frames per positive utterance) for $\calL_{frame}$.

Our models are trained with the ADAM algorithm~\cite{KingmaBa15a} with a minibatch
size of 10 utterances, an initial stepsize of $0.0001$,  for 15 epochs. 
We tune the hyperparameter $\alpha$ over the grid $\cbr{0.1, 0.5, 1, 5, 10}$
on the development set. For each $\alpha$, we monitor the model's performance on
the development set, and select the epoch that gives the lowest ER.

\begin{table}[t]
\centering
\caption{ER results of our model on the development set for different RNN architectures. 
  Here the training set size is $5000$, and we fix the number of GRU layers to be $3$.}
\label{t:result-5k}
\begin{tabular}{@{}|c|c|c|c|@{}}
\hline
& babycry & glassbreak & gunshot \\ \hline \hline
uni-directional & 0.24 & 0.06 & 0.31 \\ \hline
bi-directional & 0.18 & 0.07 & 0.26 \\ \hline
multi-resolution  & 0.13  & 0.04 & 0.20 \\ \hline
\end{tabular}
\end{table}

\begin{figure*}[t]
\centering
\begin{tabular}{@{}c@{\hspace*{0.01\linewidth}}c@{\hspace*{0\linewidth}}c@{\hspace*{0\linewidth}}c@{}}
& babycry & glassbreak & gunshot \\
\rotatebox{90}{\hspace*{7em}ER} & 
\includegraphics[width=0.32\linewidth, bb=0 20 700 650, clip]{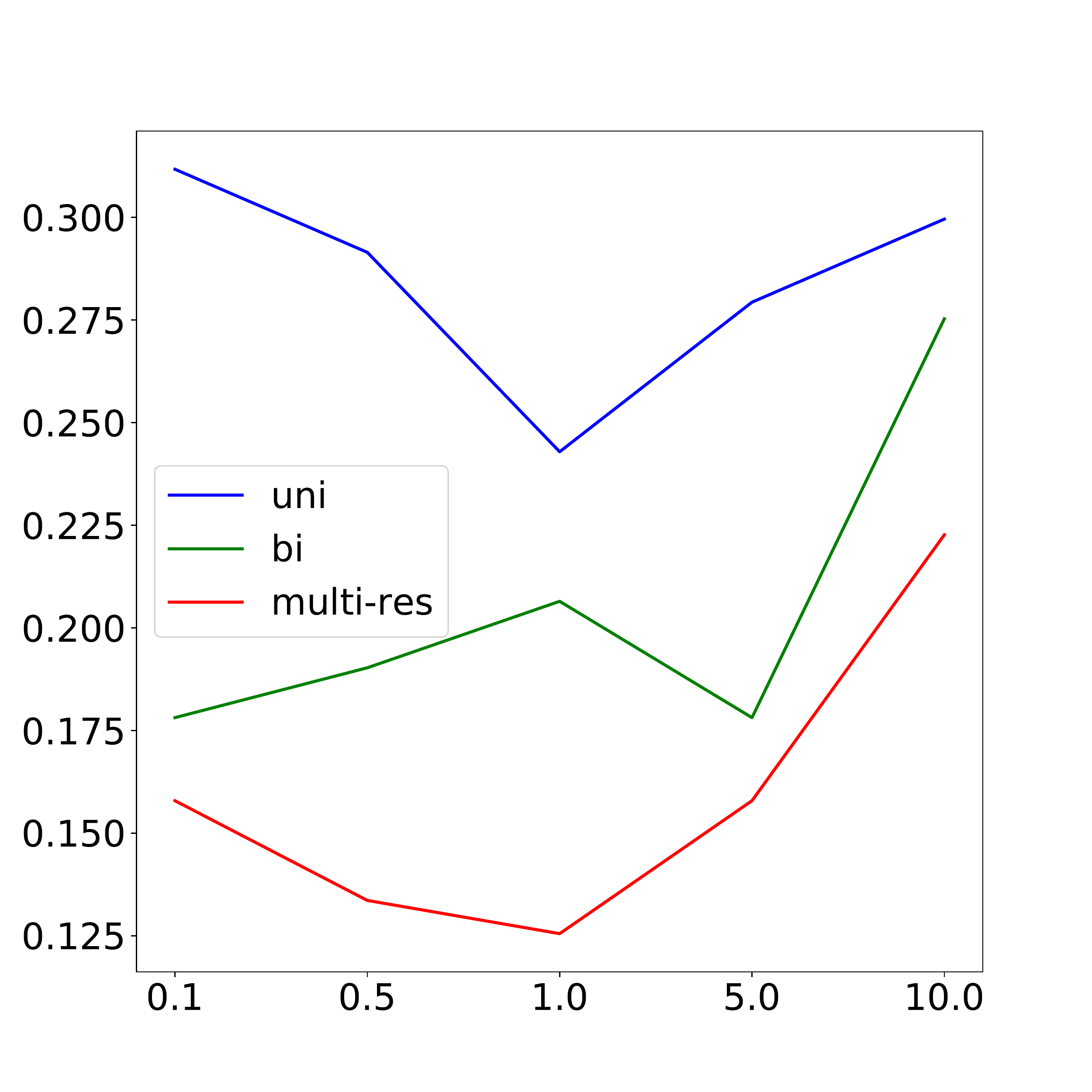} & 
\includegraphics[width=0.32\linewidth, bb=0 20 700 650, clip]{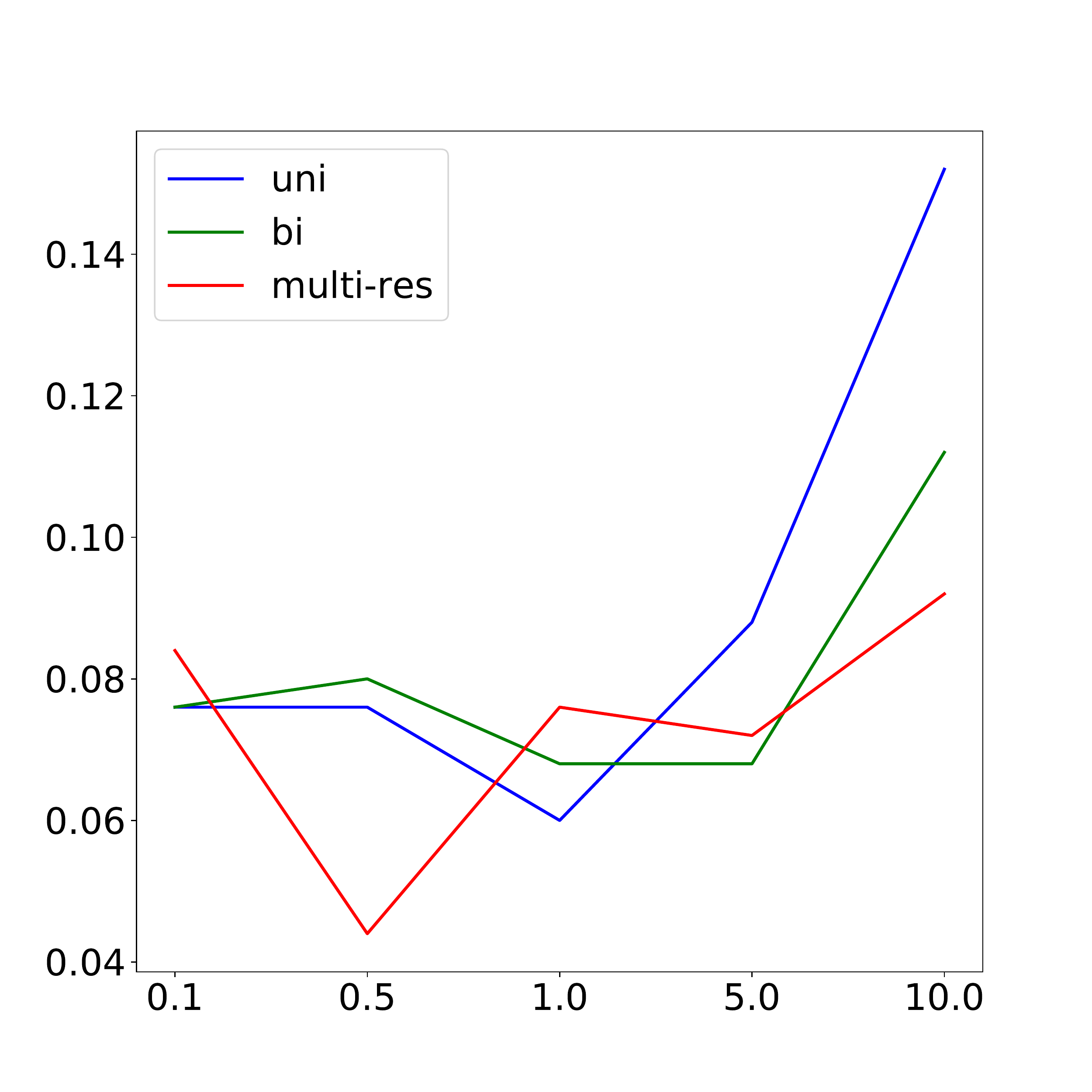} & 
\includegraphics[width=0.32\linewidth, bb=0 20 700 650, clip]{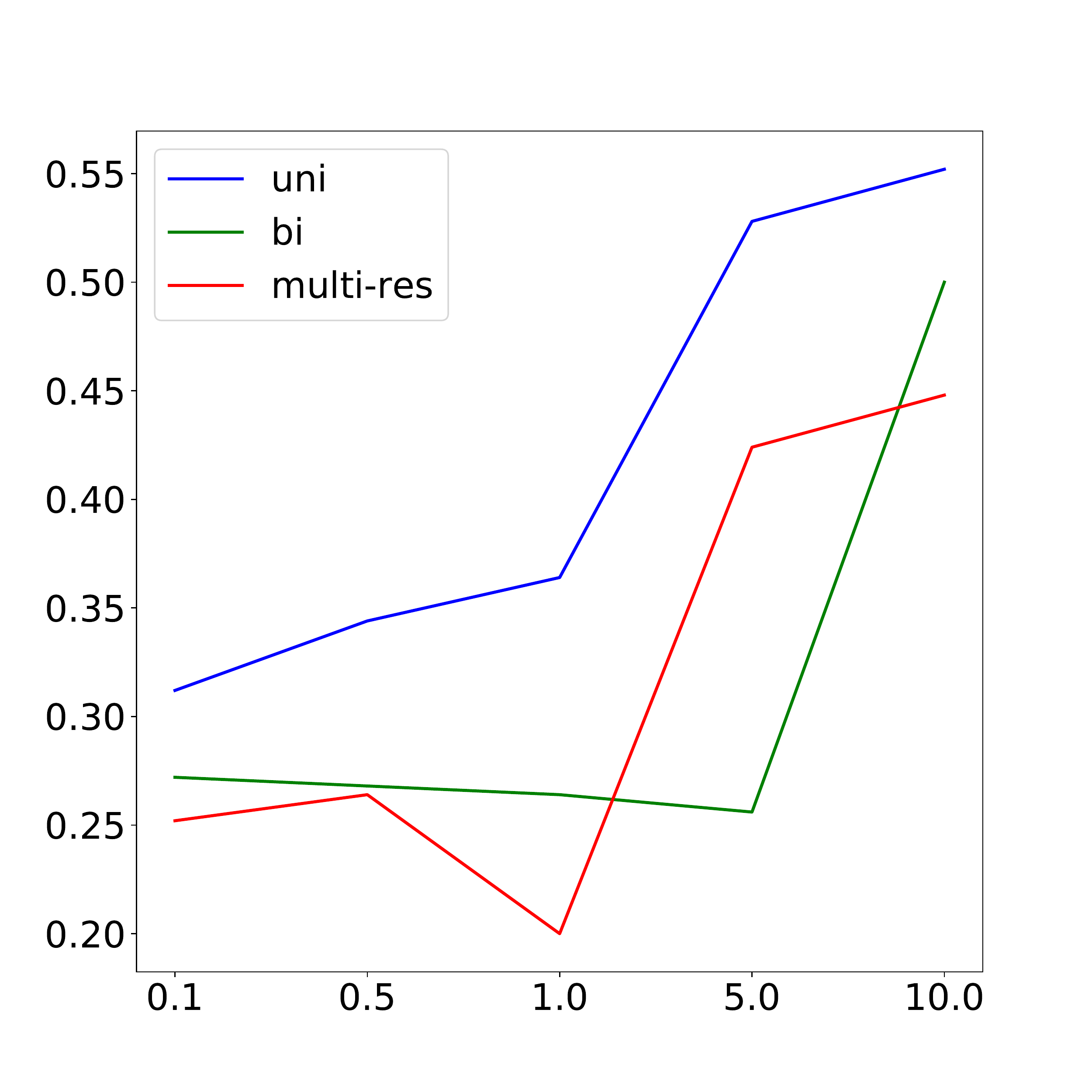} \\ [-1.5ex]
& $\alpha$ & $\alpha$ & $\alpha$
\end{tabular}
\vspace*{-1ex}
\caption{Performance of different RNN architectures for a range of
  $\alpha$. Here the training set size is $5000$.}
\label{f:alpha}
\end{figure*}

\begin{table*}[t]
\centering
\caption{Performance of our model with 15000 training samples and $4$ GRU layers.}
\label{t:result-15k}
\begin{tabular}{@{}|c|c|c|c|c|c|c|c|c|c|@{}}
\hline
& \multirow{ 2}{*}{Methods} & \multicolumn{2}{c|}{babycry}  &  \multicolumn{2}{c|}{glassbreak} &
 \multicolumn{2}{c|}{gunshot} & \multicolumn{2}{c|}{average} \\ \cline{3-10} 
& & ER & F1 (\%) & ER & F1  (\%) & ER & F1  (\%) & ER & F1  (\%) \\ \hline
  \hline
\multirow{ 4}{*}{\caja{c}{c}{Development \\ set}}
& Ours & 0.11  & 94.3 & 0.04 & 97.8  & 0.18 & 90.6 & 0.11 & 94.2 \\ \cline{2-10} 
& DCASE Baseline  
& \cellcolor{light-gray} 
& \cellcolor{light-gray} 
& \cellcolor{light-gray}
& \cellcolor{light-gray} 
& \cellcolor{light-gray}
& \cellcolor{light-gray} 
& 0.53 & 72.7 \\ \cline{2-10} 
& DCASE 1st place~\cite{Lim_17a}  
& \cellcolor{light-gray} 
& \cellcolor{light-gray} 
& \cellcolor{light-gray}
& \cellcolor{light-gray} 
& \cellcolor{light-gray}
& \cellcolor{light-gray} 
& 0.07 & 96.3 \\ \cline{2-10} 
& DCASE 2nd place~\cite{Cakir_17a}  
& \cellcolor{light-gray} 
& \cellcolor{light-gray} 
& \cellcolor{light-gray}
& \cellcolor{light-gray} 
& \cellcolor{light-gray}
& \cellcolor{light-gray} 
& 0.14 & 92.9 \\
\hline \hline
\multirow{ 4}{*}{\caja{c}{c}{Evaluation\\ set}}
& Ours & 0.26 & 86.5  & 0.16 & 92.1 & 0.18 & 91.1 & 0.20 & 89.9 \\ \cline{2-10} 
& DCASE Baseline 
& \cellcolor{light-gray} 
& \cellcolor{light-gray} 
& \cellcolor{light-gray}
& \cellcolor{light-gray} 
& \cellcolor{light-gray}
& \cellcolor{light-gray} 
& 0.64 & 64.1 \\ \cline{2-10} 
& DCASE 1st place  
& \cellcolor{light-gray} 
& \cellcolor{light-gray} 
& \cellcolor{light-gray}
& \cellcolor{light-gray} 
& \cellcolor{light-gray}
& \cellcolor{light-gray} 
& 0.13 & 93.1 \\ \cline{2-10} 
& DCASE 2nd place 
& \cellcolor{light-gray} 
& \cellcolor{light-gray} 
& \cellcolor{light-gray}
& \cellcolor{light-gray} 
& \cellcolor{light-gray}
& \cellcolor{light-gray} 
& 0.17 & 91.0 \\
\hline
\end{tabular}
\end{table*}

\subsubsection{Effect of RNN architectures}
\label{s:rnn}

We explore the effect of RNN architectures for the frame feature
transformation. 
We test 3 layers of uni-directional, bi-directional, and multi-resolution
RNNs described in Section~\ref{s:multires}  for $f(\X)$. The specific RNN cell we use
is the standard gated recurrent units~\cite{Cho_14a},  with 256 units in each direction. 
We observe that bi-directional RNNs tend to outperform uni-directional
RNNs, and on top of that, the multi-resolution architecture brings further
improvements on all events types.

\subsubsection{Effect of the $\alpha$ parameter}
\label{s:alpha}

In Figure~\ref{f:alpha}, we plot the performance of different RNN
architectures at different values of trade-off parameter $\alpha$. We
observe that there exists a wide range of $\alpha$ for which the model achieves
good performance. And for all three events, the optimal $\alpha$ is close to
$1$, placing equal weight on the utterance loss and frame loss.

\subsection{Training with 15K samples}
\label{s:final-expt}

For each type of event, we then increase the training set to 15000 utterances,
7500 of which contain the event. We use $4$ GRU layers in our
multi-resolution architecture, and set $\alpha=1.0$. 
Training stops after $10$ epochs and we perform early stopping on the development
set as before. 

The results of our method, in terms of both ER and F1-score, are given in
Table~\ref{t:result-15k}. With the larger training set and deeper
architecture, our development set ER performance is further improved on babycry and
gunshort; the averge ER of $0.11$ is only worse than the first place's
result of $0.07$ among all challenge participants.

\section{Conclusion}
\label{s:conclusion}

We have proposed a new recurrent model for rare sound events detection, which achieves competitive performance on Task 2 of the DCASE 2017 challenge. The model is simple in that instead of heuristically aggregating frame-level predictions, it is trained to directly make the utterance-level prediction, with an objective that combines losses at both levels through an attention mechanism.
To  be robust to the variations in the time axis, we also propose a multi-resolution feature extraction architecture that improves over standard bi-directional RNNs. 
Our  model can be trained efficiently in an end-to-end fashion, and thus can scale up to larger datasets and potentially to the simultaneous detection of multiple events.

\section{Acknowledgements}
The authors would like to thank Ming Sun and Hao Tang for useful
discussions, and the anonymous reviewers for constructive feedback.

\bibliographystyle{IEEEtran}
\bibliography{macp,macp-xref}

\end{document}